\documentclass[runningheads, envcountsame, a4paper]{llncs}

\usepackage{enumerate}
\usepackage{graphicx}
\usepackage[table,xcdraw]{xcolor}
\usepackage[margin=1.00in]{geometry}

\usepackage{multirow}

\newcommand*\samethanks[1][\value{footnote}]{\footnotemark[#1]}

\begin{document}

\date{}

\title{{\it Ab initio} Prediction of RNA Nucleotide Interactions with Backbone $k$-Tree Model}

\author{Liang Ding\inst{1}\thanks{To whom correspondence should be addressed.}\and Xingran Xue\inst{1}\and Sal LaMarca\inst{1}\and Mohammad Mohebbi\inst{1}\and \\Abdul Samad\inst{4}\and Russell L. Malmberg\inst{2,3}\and Liming Cai\inst{1,2}\samethanks[1]}
\institute{{\ \inst{1}Department of Computer Science, \ \inst{2}Institute of Bioinformatics \\and \  \inst{3}Department of Plant Biology,
University of Georgia, GA 30602, USA, \\
\email{\{lding, xrxue, slamarca, mohebbi\}@uga.edu, russell@plantbio.uga.edu, cai@cs.uga.edu}\\
\inst{4} Department of Computer Science, BUITEMS, Pakistan.}\\
\email{abdul.samad1@buitms.edu}}
\titlerunning{{\it Ab initio} Prediction of RNA Nucleotide Interactions with Backbone $k$-Tree Model}
\authorrunning{Ding, L. et al.}
\tocauthor{Ding, L. et al.}

\maketitle

\begin{abstract}
Given the importance of non-coding RNAs to cellular regulatory functions and rapid growth of RNA transcripts, computational prediction of RNA tertiary structure remains highly demanded yet significantly challenging. Even for a short RNA sequence, the space of tertiary conformations is immense; existing methods to identify native-like conformations mostly resort to random sampling of conformations to gain computational feasibility. However native conformations may not be examined and prediction accuracy may be compromised due to sampling. In particular, 
the state-of-the-art methods have yet to deliver the desired prediction performance for RNAs of length beyond 50.

This paper presents the work to tackle a key step in the RNA tertiary structure prediction problem, the prediction of the nucleotide interactions that constitute the desired tertiary structure.
The research is established upon a novel graph model, called {\it backbone $k$-tree}, to markably constrain 
nucleotide interaction relationships in RNA tertiary structure. It is shown that the new model makes it possible to efficiently predict the optimal set of nucleotide interactions from the query sequence, including the interactions in all recently revealed families. 
Evident by the preliminary results, the new method can predict with a high accuracy the nucleotide interactions that constitute the tertiary structure of the query sequence, thus providing a viable solution towards {\it ab initio} prediction of  RNA tertiary structure.
\end{abstract}

\section{Introduction}
In the past decade, there have been many revelations of the importance of non-coding RNAs to cellular regulatory functions and thus a growing interest in computational prediction of RNA tertiary structure \cite{LaingAndSchlick2010}, \cite{LeontisAndWesthof2012}. Nevertheless, RNA tertiary structure prediction from a single RNA sequence is a significant challenge. One major unresolved issue is in the immense  space of tertiary conformations even for a short RNA sequence. Existing methods usually employ random sampling algorithms for computation feasibility, which assemble sampled tertiary motifs into native-like structures \cite{DasAndBaker2007}, \cite{DingEtAl2008}, \cite{JonikasEtAl2009}, \cite{ParisienAndMajor2008}, \cite{PopendalEtAl2012}, \cite{SharmaEtAl2008}. To reduce the chance to miss native structures, the assembly algorithms have mostly been guided with constraining structural models. For example, MC-Fold/MC-Sym \cite{ParisienAndMajor2008} assumes the tertiary structure consists of 4-nt cyclic tertiary motifs constructible from the predicted secondary structure. Rosetta \cite{DasAndBaker2007,DasEtAl2010} {\it de novo} assembles tertiary structure from a database of 3-nt tertiary fragments. Other methods follow samplings that preserve the secondary structure \cite{BidaAndMaher2012}, \cite{PopendalEtAl2012,ReinharzEtAl2013} or intervention from human experts \cite{JossinetEtAl2010}, \cite{MartinezEtAl2008}. However, these constraining models do not necessarily ensure that native conformations are 
examined. In particular, the state-of-the-art methods have yet to deliver the desired prediction accuracy for RNA sequences of lengths beyond 50 \cite{LaingAndSchlick2010}.

In this work, we introduce a novel method to predict nucleotide interactions from sequences as a key step toward accurate {\it ab initio} prediction of tertiary  structure. 
Accurate knowledge of the nucleotide interactions is crucial to predicting the tertiary structure of an RNA and subsequently predicting its functional roles. To predict nucleotide interactions, our method is guided by a novel graph model called a {\it backbone $k$-tree}, for small integer $k$, to globally constrain the nucleotide interaction relationships (NIRs) that constitute the tertiary structure.  In such a $k$-tree graph, nucleotides are organized into groups of size $k+1$, such that NIRs are permitted only for nucleotides belonging to the same group and groups are connected to each other with a tree topology (see section 2). This model was inspired by our recent discovery of the small treewidth of the NIR graphs for more than 3,500 RNA chains extracted from 1,984 resolved RNAs (Figure 1). We have been able to develop dynamic programming algorithms with $O(n^{k+1})$ time and space complexities, efficient for small  $k$, to compute the optimal backbone $k$-tree spanning over the nucleotides on the query sequence, given a scoring function \cite{DingEtAl2014,DingEtAl2014a}


\begin{figure}[!ht]
\centering
\includegraphics[scale=0.40]{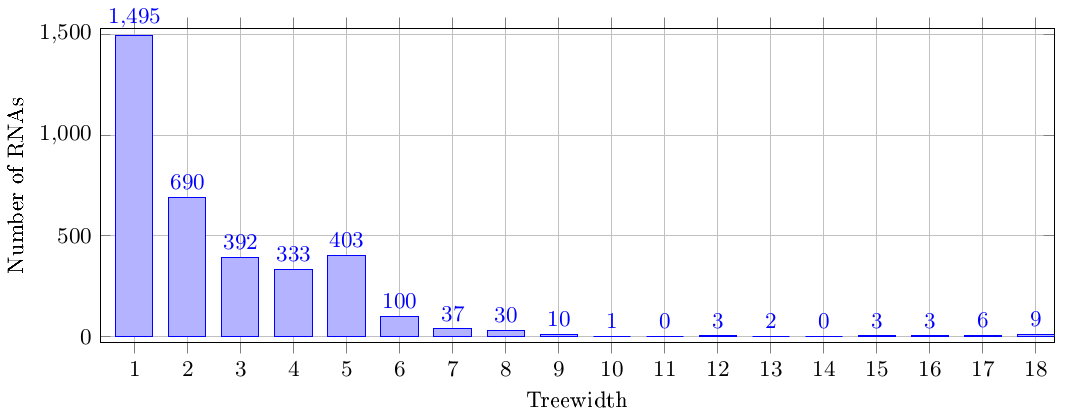}
\caption{Treewidth distribution of NIR graphs of more than 3,500 chains deriving from 1,984 resolved RNA tertiary structures in the RNA Structure Atlas \cite{SarverEtAl2008}. The RNAs with treewidth larger than 18 are omitted due to their very small number. These treewidths are actually upper bounds computed by a program \cite{BoalaenderAndKoster2010}; it is likely that the exact treewidths of the NIR graphs may actually be smaller.}
\label{tw}
\end{figure}

To ensure that the computed optimal $k$-tree can actually yield the set of nucleotide interactions that constitutes the native tertiary structure, our method defines the scoring function over detailed patterns of nucleotide interactions within every group of $k+1$ nucleotides. We consider nucleotide interactions from the established geometric nomenclatures \cite{LeontisAndWesthof2001} and nucleotide interaction families \cite{LeontisEtAl2002}, \cite{StombaughEtAl2009}, \cite{ZirbelEtAl2009}, including base-base, base-phosphate, and base-ribose as well as base-stacking interactions. To test our method, we adopted an improved 3-tree model and pre-computed candidates of interaction patterns for every group of 4 nucleotides, by searching through RNA Structure Atlas \cite{SarverEtAl2008}; this contains annotated atom-level nucleotide interactions for nearly 3,000 resolved tertiary structures. 
We trained artificial neural networks (ANNs) to compute the confidence of every given nucleotide interaction and the confidence of every admissible nucleotide interaction pattern for every group of 4 given nucleotides. We filtered out unlikely interaction patterns and kept only those with high confidences. With this 3-tree model, our algorithm efficiently predicts an optimal set of nucleotide interactions from the query sequence within computational time $O(c^5Mn^3)$, where $M$ is a constant and $c\leq 20$ is the maximum number of candidate interaction patterns for one group of 4 nucleotides. We have implemented the algorithm into a program called BkTree, which may use known or predicted canonical (i.e., {\it cis} Watson-Crick) base pairs on the query sequence.

To evaluate our method for nucleotide interaction prediction, we tested BkTree on a benchmark set of 43 high resolution RNAs, which had been used to survey a number of state-of-the-art tertiary structure prediction methods \cite{LaingAndSchlick2010}. The resolved, atom-level interactions were extracted with FR3D \cite{SarverEtAl2008}. BkTree performed impressively well across the set of tested RNAs (Table~\ref{withBbsTable}), achieving the averaged sensitivity, PPV, and MCC values of 0.86, 0.78, and 0.82, respectively (discounting the input canonical base pairs). In comparison with previous programs MC \cite{ParisienAndMajor2008}, Rosetta \cite{DasAndBaker2007}, and NAST \cite{JonikasEtAl2009} that all assumed the secondary structure as a part of the input \cite{LaingAndSchlick2010}, it is clear that BkTree outperformed the other three programs in the MCC measure on this set of benchhmark RNAs (Table~\ref{avg4methods}, Figure~\ref{laingMCCCompare}).  In particular, on the four representative RNAs that contain typical helices and junctions \cite{LaingAndSchlick2010}, BkTree gave the best performance on all but one RNA, for which BkTree acquired a higher sensitivity value but lower PPV than the MC program, resulting in a slightly lower MCC value (Table~\ref{compare_Laing}).

To evaluate the significance of our method to 3D conformation prediction, we used the program MC-Sym to model 3D conformations from the interactions predicted by BkTree and calculated RMSDs against the resolved structures. Since MC-Sym requires secondary structure for 3D conformation modeling, we identified 30 RNAs from the benchmark set for which their secondary structures are covered by the BkTree-predicted nucleotide interactions together with its input canonical base pairs. 
For the 4 representative RNAs listed in Table~\ref{compare_Laing}, BkTree outperforms MC and Rosetta on 3 of them. 





\section{Model and Methods}

In this work, we consider all known types of 
nucleotide interactions of atomic-resolution \cite{LeontisAndWesthof2001}, \cite{LeontisEtAl2002}, \cite{ZirbelEtAl2009}.
In particular, with the base triangle model consisting of Watson-Crick (W), Hoogsteen (H), and sugar (S) edges, base-base interactions has been fully characterized into rich 12 geometric types and 18 interaction families  \cite{LeontisAndWesthof2001}, \cite{LeontisEtAl2002}, according to involved edges, {\it cis} or {\it trans}, and parallel or anti-parallel, observed in crystal structures. For example the cWW family contains, in addition to the canonical (i.e., {\it cis} Watson-Crick) base pairs, many non-canonical base-base interactions through W edges. More recently, classifications of nucleotide interactions have been extended to base-backbone interactions. There are 10 families identified for base-phosphate interactions based on the position of the interacting hydrogen atom in the base \cite{ZirbelEtAl2009}. Similarly, 9 additional families have been identified for base-ribose interactions \cite{ZirbelEtAl2011}. A few base stacking interactions have also been classified. 
Table 1 summarizes these classes of nucleotide interactions, which also includes the backbone interaction between two neighboring nucleotides.

\begin{table}[!ht]
\centering
\caption{Categories, types and families of RNA nucleotide interactions, mostly summarized from works \cite{LeontisAndWesthof2001}, \cite{LeontisEtAl2002}, \cite{ZirbelEtAl2009,ZirbelEtAl2011}. It also includes the  phosphodiester interaction between two neighboring nucleotides.}
\begin{tabular}{|cl|c|}
\hline
\multicolumn{1}{|c|}{Categories} & \multicolumn{1}{c|}{Types (Interaction Families)} & Number \\ \hline\hline
\multicolumn{1}{|c|}{\multirow{2}{*}{Base pairs}} & cWW, tWW, cWH, tWH, cHW, tHW, cWS, tWS, cSW, & \multirow{2}{*}{18} \\
\multicolumn{1}{|c|}{} & tSW, cHH, tHH, cHS, tHS, cSH, tSH, cSS, tSS &  \\ \hline
\multicolumn{1}{|c|}{Base-phosphates} & 0BPh, 1BPh, 2BPh, 3BPh, 4BPh, 5BPh, 6BPh, 7BPh, 8BPh, 9BPh & 10 \\ \hline
\multicolumn{1}{|c|}{Base-riboses} & 0BR, 1BR, 2BR, 3BR, 4BR, 5BR, 6BR, 7BR, 9BR & 9 \\ \hline
\multicolumn{1}{|c|}{Bases stackings} & s35, s53, s33, s55 & 4 \\ \hline
\multicolumn{1}{|c|}{Backbone-backbone} & phosphodiester & 1 \\ \hline
\end{tabular}
\label{all_inters}
\end{table}
 
\subsection{Backbone $k$-Tree Model}

Let the query RNA sequence be $S=S_1 S_2, \dots S_n$, where $S_i \in \{{\tt A, C, G, U}\}$, for $1\leq i \leq n$. We denote an interaction between 
the $i$th and $j$th nucleotides, where $i<j$, with triple $\langle{S_i}^{(i)}, {S_j}^{(j)}, t\rangle$, for some interaction type $t$ shown in Table~\ref{all_inters}. Note that there are possibly two or more  simultaneous interactions between the two nucleotides.

Given the native tertiary structure of the sequence $S$, we 
model the {\it nucleotide interaction relationship}s (NIRs) within the tertiary structure with a  graph $G=(V, E)$, where $V=\{ {S_i}^{(i)}: 1\leq i \leq n\}$, such that $({S_i}^{(i)}, {S_j}^{(j)})$ is an edge in $E$ if and only if $i\not= j$ and $\langle{S_i}^{(i)}, {S_j}^{(j)}, t\rangle$ is an interaction for some $t$. 
We call $G$ {\it the NIR graph} of the sequence with the given structure. Because every two consecutive nucleotides are connected with the phosphodiester bond, every NIR graph of $n$ vertices contains all edges $({S_i}^{(i)}, S_{i+1}^{(i+1)})$, for $1\leq i \leq n-1$. These edges are called {\it backbone edges}.

In our recent investigation \cite{DingEtAl2014}, we constructed NIR graphs for all RNAs whose tertiary structures were known from RNA Structure Atlas \cite{ReinharzEtAl2013}. We discovered that an overwhelming majority of these RNAs are of small treewidths (Figure~\ref{tw}). Treewidth is a graph metric, which intuitively indicates how much a graph is tree-like. If a graph has treewidth bounded by $k$, any clique obtained by deleting vertices and edges and contracting edges of the graph can contain at most $k+1$ vertices \cite{ArnborgEtAl1990}. Thus the distribution of treewidths suggest that NIRs in the RNA tertiary structures are in general not arbitrarily complex. 

The concept of treewidth originated from the algorithmic graph theory. It is closely related to, and may be better explained with the notion of {\it $k$-tree}, which is central to this work. 

\begin{definition} \rm \cite{Patil1986}
\label{k-tree}
Let integer $k\geq 1$. The class of {\it $k$-trees} are graphs defined by the following inductive steps:
\begin{enumerate}
\item A $k$-tree of $k+1$ vertices is a clique of $k+1$ vertices;
\item A $k$-tree of $n$ vertices, for $n>k+1$, is a graph consisting of a $k$-tree $G$ of $n-1$ vertices and a vertex $v$, which does not occur in $G$, such that $v$ forms a $(k+1)$-clique with some $k$-clique already in $G$.
\end{enumerate}
\end{definition}

\begin{figure}[!ht]
\centering
\includegraphics[width=\textwidth]{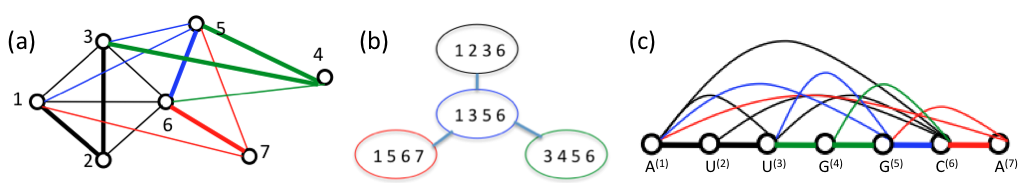}
\caption{(a) 3-tree of 7 vertices by Definition~\ref{k-tree}, with the order of forming the four 4-cliques: with initial clique $\{1, 2, 3, 6\}$ (black edges), vertex $5$ and blue edges added, then vertex $7$ and red edges added, and finally vertex 4 and green edges added. (b) Illustration of the graph of (a) with a tree-topology connecting the four 4-cliques. (c) A backbone 3-tree for sequence {\tt AUUGGCA}, of the same topology as shown in (a); backbone edges are in bold.}
\label{fig_k_tree}
\end{figure}

Figure \ref{fig_k_tree} shows a 3-tree with seven vertices in (a) and illustrates it in (b) with a tree-topology that connects the four 4-cliques in the graph. 

By \cite{vanLeeuwen1990}, for any $k \geq 1$, a graph is of treewidth $\leq k$ if and only if it is a subgraph of a $k$-tree. Therefore, NIR graphs for an overwhelming majority of known RNA tertiary structures are constrained in topology by $k$-trees, for small values of $k$. Because technically, every graph of treewidth bounded by $k$ can be augmented  with additional edges into a $k$-tree, we adopt such $k$-trees as the model for NIRs of the RNA tertiary structure.

\begin{definition} \rm Let $k \geq 1$ be an integer. The {\it backbone $k$-tree} for an RNA sequence is an augmented NIR graph of the sequence, which is a $k$-tree.
\end{definition}

Figure 1(c) shows a backbone $3$-tree for sequence {\tt AUUGGCA}. Note that backbone $k$-trees differ  from general $k$-trees in that a backbone $k$-tree has to the designated Hamiltonian path (consisting of all the backbone edges).

With the backbone $k$-tree model, in order to predict the set $I$ of nucleotide interactions from the query sequence, we propose to identify a backbone $k$-tree $G=(V, E)$ such that
\[ (S_i^{(i)}, S_j^{(j)}) \in E \mbox{ if and only if } \exists \, t\, \mbox{ } \langle S_i^{(i)}, S_j^{(j)}, t\rangle  \in I\]

To ensure the identified $G$ actually corresponds to the set of interactions that constitute the native structure of the query sequence, we need to quantify nucleotide interactions for combinatorial optimization of such a backbone $k$-tree $G$, as explained in the subsequent sections.

\subsection{Quantification of Nucleotide Interactions}

\begin{definition} {\rm Let $q$ be a $(k+1)$-clique in a backbone $k$-tree of query sequence $S$. 
An {\it interaction pattern} (ip) for clique $q$ is a set $P_q$ of interactions for the nucleotides in $q$ such that for every interaction $\langle {S_i}^{(i)}, {S_j}^{(j)}, t\rangle$ in $P_q$, both nucleotides ${S_i^{(i)}}$ and ${S_j}^{(j)}$ are in clique $q$. 
}
\end{definition}

Given an ip $P_q$ for clique $q$, we define the {\it induced subgraph by $P_p$}, denoted with $B_{P_q} = (q, E_{B_{P_q}})$ to be a subgraph of $q$ such that edge $(S_i^{(i)}, S_j^{(j)}) \in E_{B_{P_q}}$ only if interaction $\langle S_i^{(i)}, S_j^{(j)}, t\rangle \in P_q$ for some $t$.

\begin{definition} {\rm Let $q$ be a $(k+1)$-clique in the in a backbone $k$-tree of query sequence $S$. The {\it confidence} of a given ip $P_q$ for clique $q$ is defined as
\begin{eqnarray}
\label{confidence}
f(q, P_q, S) = \sum_{\langle S_i^{(i)},S_j^{(j)}, t \rangle \in P_q} c_{q, B_{P_q}, t}^{(i,j)} 
\end{eqnarray}
where $c_{q, B_{P_q}, t}^{(i,j)}$ is the {\it confidence} of interaction 
$\langle S_i^{(i)}, S_j^{(j)}, t\rangle$ given $q$ and subgraph $B_{P_q}$ induced by $P_q$.}
\end{definition}

In the Section~\ref{section-algo}, we will introduce artificial neural networks (ANNs) to compute confidence $c_{q, B_{P_q}, t}^{(i,j)}$. 

For every clique $q$, with ${\cal Q}(q)$, we denote the finite set of all ips for $q$. In the practical application, we may only include those ips in ${\cal Q}(q)$ which have ``high'' confidences (e.g., above certain threshold). Let $I$ be a set of interactions. By notation $I|_q$, we mean the maximal size subset of $I$ that is an ip for $q$.

\begin{definition} \rm Let $k$ be any fixed integer $\geq 2$. The {\it nucleotide interaction prediction} problem NIP$(k)$ is, given an input query sequence $S$, to identify a backbone $k$-tree $G^*=(V, E^*)$ as well as a set $I^*$ of nucleotide interactions that constitutes the tertiary structure of $S$, such that 
 every interaction $\langle S_i^{(i)},S_j^{(j)}, t \rangle \in I^*$ implies edge  $(S_i^{(i)},S_j^{(j)}) \in E^*$ and 
\begin{equation}
\label{optimization}
(I^*, G^*) = \arg \, \max\limits_{(I, G)} \,\{ \sum\limits_{q \mbox{ in } G, \, I|_q \in  {\cal Q}(q)}  f(q, I|_q, S)\}
\end{equation}
\end{definition}

\subsection{Overview of the Method}\label{section-overview}

Our method consists of three major components to solve the NIP$(k)$ problem, for any fixed $k\geq 2$. The first component is data repositories including NIPDB and NIPCCTable. NIPDB is a database of all possible interaction patterns (ips) for every $(k+1)$-clique, which was established by searching through the RNA Structure Atlas \cite{SarverEtAl2008}. For every such clique, its ips in NIPDB are extracted and ranked when the query sequence is preprocessed. NIPCCTable is a matrix for compatibility between every pair of ips for two cliques that share all but one nucleotide. The compatibility is checked by the dynamic programming algorithm computing the NIP$(k)$ problem.

The second component is a set of artificial neural networks (ANNs) to compute confidence for any given interaction type $t$ between any two given nucleotides $S_i^{(i)}$ and $S_j^{(j)}$ on the query sequence. The computed confidences for interactions are then used to compute confidence of an ip for every $(k+1)$-clique, as formulated in equation~(\ref{confidence}). For every such clique $q$, all ips of $q$ obtained from database NIPDB are ranked according to their confidence values. Often the number of ips with significant confidence values is small, e.g., $\leq 20$; ips of significant scores are included as ip candidates into the set ${\cal Q}(q)$ for $q$.  The detailed construction of the ANNs will be described in the next section. 



The third component is a dynamic programming algorithm solving the NIP$(k)$ problem, using the prepared data and preprocessing results from the first two components. From the input query sequence, the algorithm produces a backbone $k$-tree $G^*$ as well as a set $I^*$ of nucleotide interactions, maximizing the aggregate confidence value across all $(k+1)$-cliques in $G^*$ (see equations (\ref{optimization}) and (\ref{confidence})). The relationship between $G^*$ and $I^*$ is that, for every $(k+1)$-clique $q$ in the $k$-tree $G^*$,  there is a maximal subset of nucleotide interactions $I|_q \subseteq I^*$ being an ip for $q$, such that $I^* = \bigcup_{q \mbox{ in } G^*} I^*|_q$.  The next section describes the details of the  dynamic programming algorithm.

\section{Algorithms}
\label{section-algo}

\subsection{ANNs for Computing Interaction Confidence}

Let the query sequence $S=S_1S_2\dots S_n$ of $n$ nucleotides, where $S_i \in \{{\tt A, C, G, U}\}$, for $1\leq i \leq n$. Technically we considered all $(k+1)$-cliques formed by $k+1$ vertices $\{S_{h_0}^{(h_0)}, S_{h_1}^{(h_1)}, \dots, S_{h_k}^{(h_k)}\}$, where $1\leq h_0 < h_1 < \dots < h_k \leq n$. Let $q=(V, E)$ be such a clique and $B_q = (V, E_{B_q})$,  where $E_{B_q} \subseteq E$, be any subgraph of $q$. For every edge $(S_i^{(i)},S_j^{(j)})\in E_{B_q}$ and every possible interaction $\langle S_i^{(i)},S_j^{(j)}, t\rangle$ of type $t$, we constructed an ANN ${\cal N}_{q, B_{q},t}^{(i,j)}$ to calculate {\it confidence} $c_{q, B_{q},t}^{(i,j)}$ that interaction
$\langle {S_i}^{(i)}, {S_j}^{(j)}, t\rangle$ occurs in the subgraph $B_q$ of clique $q$.

Each ANN ${\cal N}_{q, B_{q},t}^{(i,j)}$ consists of an input layer, a hidden layer, and an output layer. The output layer is a single unit depicting a confidence value for interaction $\langle {S_i}^{(i)}, {S_j}^{(j)}, t\rangle$. The input layer consists of input units representing the selected global and local features shown in Table \ref{featureTable}. The features included the sequence length and the distance between the involved nucleotides as well as neighboring nucleotide types. In addition, we included the information of {\it assumed} canonical base pairs\footnote{These are known or predicted Watson-Crick and wobble base  pairs. Note that they do not necessarily constitute all information about the secondary structure.} within the query sequence. The complete list of features selected for the trainings are given in the Table \ref{featureTable}.

\begin{table}[!ht]
\caption{Features selected from a given $(k+1)$-clique $q$ and given subgraph $B_q$ of $q$ for training ANN ${\cal N}_{q, B_{P_q},t}^{(i,j)}$. CBP is an abbreviation for canonical base pair. A {\it Component} (Cp) is defined as the maximal subsequence consisting of two or more nucleotides each involved in a CBP.}
\centering
\begin{tabular}{|c|c|p{10cm}|}
\hline
Feature & Value & \multicolumn{1}{c|}{Comments} \\ \hline\hline
Seq. length & An integer & Length of a training sequence containing $q$. \\ \hline
Distances & $k$ integers & Distances between every two nucleotides in the sequential order in  $q$. \\ \hline
\multirow{3}{*}{Number of Cps} & \multirow{3}{*}{$k$ integers} & Number (one of $\{0, 1, 2, 3, -1\}$) of Cps on the subsequence between every two nucleotides in the sequential order. 3 means there are at least 3 Cps; $-1$ means the two nucleotides are neighboring nucleotides on the sequence. \\ \hline
\multirow{4}{*}{Neighbor nts.} & \multirow{4}{*}{$k+1$ 4-mers} & One $4$-mer (of letter {\tt A, C, G, U}) for every nucleotide in $q$, where the first two letters and the last two letter of the 4-mer indicate the two nts to the left and to the right of the nucleotide, respectively, and letter {\tt N} is used when there is no neighbor. \\ \hline
\multirow{4}{*}{Neighbor CBPs} & \multirow{4}{*}{$k+1$ 4-mers} & One 4-mer (of binary bits) for every nucleotide in $q$, where the first two bits and the last two bits of the 4-mer indicate the two nts to the left and to the right of the nucleotide are involved in CBPs, respectively, and letter {\tt N} is used when there is no neighbor. \\ \hline
\multirow{4}{*}{Edge properties} & \multirow{4}{*}{up to $\frac{k(k+1)}{2}$ integers} & For every edge  in the subgraph $B_q$ of $q$, value 0 indicates both nts are involved in a CBP; -1 (resp. +1) indicates exclusively left (resp. right) nt is involved in a CBP; 2 indicates either is near a CBP; and -2 indicates both are far away (distant beyond 3 nts) from a CBP. \\ \hline
\end{tabular}
\label{featureTable}
\end{table}

We adopted conventional methods to construct and train the ANNs \cite{Mitchell1997}, typically the technique of back-propagation with gradient descent, using a fixed-size network. This is based on the calculation of the error by taking the first derivatives of half the Euclidean distance between the output and target and back-propagating it towards the input layer, over the whole training set. Each weight is then updated according to the error contribution of each unit, the error of each output unit and a learning rate. The logistic sigmoid was used as the activation functions for each unit. The updating is repeated until the training error converges to a minimum or the cross-validation error starts to rise, due to over-fitting.  The learning rate $0.03$ was the value that yielded the best results for a subset of $895$ RNAs from RNA Structure Atlas.

The trained ANNs can be applied to compute confidence for interaction patterns. In particular, given a $(k+1)$-clique $q =\{S_{h_1}^{(h_1)}, S_{h_2}^{(h_2)}, \dots, S_{h_{k+1}}^{(h_{k+1})}\}$, $1\leq h_1  < \dots < h_{k+1} \leq n$, let $P_q$ be an ip for $q$ and let $B_{P_q}$ be the underlying graph for $P_q$, which is a subgraph of clique $q$. Then the trained ANN ${\cal N}_{q, B_{P_q},t}^{(i,j)}$ can be applied on each edge $(S_i^{(i)},S_j^{(j)}) \in E_{B_{P_q}}$ and each type $t$ to compute the confidence  score  $c_{q,B_{P_q}, t}^{(i,j)}$ for interaction $\langle {S_i}^{(i)}, {S_j}^{(j)}, t\rangle$. The confidence $f(q, P_q, S)$ of $P_q$ for $q$ is computed with the equation~(\ref{confidence}). 

Then for $q$, all the ips $P_q$'s are ranked according to their confidences $f(q, P_q, S)$, and only significant top $m$ ips are included in the candidate set ${\cal Q}(q)$. We have chosen $m \leq 20$ in the performance evaluations as our experiments results had showed that a larger $m$ could not help to improve the results.

\subsection{Algorithm for NIP$(k)$ problem}

Roughly speaking, the algorithm for NIP$(k)$ problem considers every $(k+1)$-clique, from which recursive creations of more cliques are all examined. For every newly created clique $q$, all ips from ${\cal Q}(q)$ 
are considered but eventually exactly one of them is chosen for $q$. The algorithm follows the basic process of creating $k$-tree given in Definition 1. However, because the identified $k$-tree is a backbone $k$-tree that contains all backbone edges, the process is not straightforward. We need the following notations for an introduction to the algorithmic idea. 
By {\it interval} $[i..j]$, for $i\leq j$, we mean the set of consecutive integers between $i$ and $j$, inclusive. Two intervals $[i..j]$ and $[h..l]$ are {\it non-overlapping} if either $j\leq h$ or $l\leq i$. 
Formally, let the query sequence be $S=S_1S_2\dots S_n$ and $q$ be a clique formed by $k+1$ vertices $\{S_{h_1}^{(h_1)}, S_{h_2}^{(h_2)}, \dots, S_{h_{k+1}}^{(h_{k+1})}\}$, where $1=h_0 \leq h_1 < h_2 < \dots < h_{k+1} \leq n=h_{k+2}$. Let $A$ be a set of non-overlapping intervals and $P_q \in {\cal Q}(q)$ be an ip for clique $q$.  

We define function $M(q, A, P_q, S)$ 
to be the maximum confidence of a $k$-tree constructed beginning from clique $q$, which includes all backbone edge $(S_i^{(i)}, S_{i+1}^{(i+1)})$ for integers $i$ and $i+1$ both  contained in the same interval in $A$. Then we obtain the following recurrence:

\begin{eqnarray}
M(q, A, P_q, S) &=& \max\limits_{S_x^{(x)} \in q, \, S_y^{(y)} \not \in q, \, y \in [i..j] \in A, \, p=q|^x_y \,} \nonumber \\
&& \{ \max_{P_p \in {\cal Q}(p), \, {\cal R}(B, C), {\cal P}(P_q, P_p) } \{ M(p, B, P_p, S) + M(q, C, P_q, S) + f(q, P_q, S) \} \, \}\label{dp}
\end{eqnarray}
where abbreviations $q|^x_y = q \cup \{S_y^{(y)} \}\setminus \{S_x^{(x)}\}$, ${\cal P}(P_p, P_q)$ asserts that the chosen ip $P_p$ be compatible with $P_q$,  and ${\cal R}(B, C)$ represents the choices of two sets of intervals, $B$ and $C$, which satisfy constraints 
\begin{enumerate}[(a)]
\item $\{[i..y], [y..j]\} \subseteq B$, $\{[w..x], [x..z]\} \subseteq C$, for applicable $w$ and $z$; and 
\item $B \cup C = A \cup \{[i..y], [y..j]\} \setminus \{[i..j]\}$, and $B \cap C = \emptyset$.
\end{enumerate}

Recurrence (\ref{dp}) gives an iterative process to produce a backbone $k$-tree. The intuitive idea is to create a new clique $p$ from $q$ by introducing a new nucleotide vertex $S_y^{(y)}$ to the partially constructed $k$-tree. This results in possibly two or more sub-$k$-trees, one starting from $p$ and the others from $q$ (but not including $S_y^{(y)}$). Since the two or more sub-$k$-trees will never join together again, interval sets are used to ensure backbone edges will be properly created. 
Essentially, the constructed $k$-tree corresponding to the value of function $M(q,A,P_q,S)$ contains only those backbone edges that connect the nucleotides of indexes specified in the intervals in $A$. In particular, starting from clique $q$ of $k+1$ vertices $\{S_{h_1}^{(h_1)}, S_{h_2}^{(h_2)}, \dots, S_{h_{k+1}}^{(h_{k+1})}\}$, to compute an backbone $k$-tree that contains all the backbone edges, we need to set $A = \{ [h_i..h_{i+1}]: 0\leq i \leq k+1\}$, where $h_0 =1$  and $h_{k+2} = n$.  

The confidence score of the produced $k$-tree is computed as the sum of confidence scores of ips chosen for all involved $(k+1)$-cliques. The chosen ips need to be  compatible across the cliques when they share nucleotide interactions or even just nucleotides. This is ensured by the assertion ${\cal P}(P_q, P_p)$, which checks (1) $P_q$ and $P_p$ have the same set of interactions on the edges shared by cliques $q$ and $p$ by looking up table NIPCCTable; and (2) any pattern of interactions between a single nucleotide and multiple others has to exist in the structure database.


To complete the recurrence, we need the following base case:
\[M(q, A, P_q, S) = 0 \, \, \,\mbox{  if }A = \emptyset\]
To identify the desired  backbone $k$-tree $G^*$, we maximize $M(q, A, P_q, S)$ over all starting clique $q$ and all ip $P_q \in {\cal Q}(q)$. The associated set $I^*$ of nucleotides is just the union of the chosen ips for all $(k+1)$-cliques in $G^*$.

Recurrence (\ref{dp}) naturally offers a dynamic programming solution. Function $M(q, P_q, A, S)$ can be computed by establishing a table with dimensions for $q, P_q$, and $A$. With the base cases, the table is computed bottom-up, from $A=\emptyset$, using the recurrence~(\ref{dp}).

\subsection{Improved Algorithms}
Simply implementing the above outlined algorithm would require $O(n^{k+1})$ memory space and $O(n^{k+2})$ computation time for every fixed value of $k$. Following the same idea but creating $(k+1)$-cliques from $k$-cliques instead leads 
to an improved dynamic programming algorithm to solve the NIP$(k)$ problem, with a little more sophisticated steps to navigate through $k$-cliques. The improved algorithm uses $O(n^{k})$ amount of memory space and $O(n^{k+1})$ amount of time for every fixed value of $k$ \cite{DingEtAl2014,DingEtAl2014a}. 

The efficiency can be further improved by demanding that every $(k+1)$-clique in backbone $k$-trees contains two consecutive nucleotides $S_i^{(i)}$ and $S_{i+1}^{(i+1)}$ for some $i$. That is, every interaction pattern for a $(k+1)$-clique always contains at least one backbone edge. This allows a further reduction of computation time to $O(n^k)$. Testing on the case $k=3$ has shown that the constrained backbone $3$-tree model maintains the similar capability to account for sophisticated nucleotide interactions as the ``standard'' backbone $3$-tree model. In addition the constraint may enforce the construction of the $3$-tree to follow backbone edges, providing more controls on the $3$-tree construction. Finally, the constraint also significantly reduced the number of cases that the ANNs need to consider in their construction.

\subsection{Implementation}
The NIPDB database construction was implemented by Python, where Prody package \cite{BakanEtAl2011} was adopted to search RNA Structure Atlas. Afterward, NIPCCTable, the matrix for ip consistence and compatibility was developed using Python. Training and building of ANNs were realized with WEKA package \cite{MarkHallEtAl2008}. Finally, confidences of ips admissible for every clique $(k+1)$-clique in the query sequence was computed by programs in Python.

We implemented in C++ the dynamic programming algorithm into a program called BkTree. We ran the evaluation tests on a Red Hat 4.8.2-7 server with 4 Intel Quad core X5550 Xeon Processors, 2.66GHz 8M Cache and 70GB Memory. 

\section{Performance Evaluation}
\subsection{Test Data}\label{sectionTestData}
We implemented our method in the program BkTree. We evaluated our method through testing BkTree on a list of 43 RNAs
of high resolution structure data, which had been used as a benchmark set to evaluate a number of state-of-the-art tertiary structure prediction methods in the survey \cite{LaingAndSchlick2010}. 18 of the RNA sequences are of length $\geq 50$. In developing the ANNs for computing interaction confidences, 7 of these RNAs were not included in the training data.

\begin{table}[!ht]
\centering
\caption{Nucleotide interaction prediction results by BkTree on the benchmark set used in the survey \cite{LaingAndSchlick2010}. The number of canonical base pairs (CPBs) and number of non-canonical interactions (NCIs) are listed. The sensitivity (STY), PPV and MCC were calculated, excluding the canonical bases pairs used as a part of the input. The data of the 7 RNAs not used for training ANNs are displayed with the bold font.}
\label{withBbsTable}
\begin{tabular}{|c|c|c|c|c|c|c|l|}
\hline
PDB ID& Length & \# CBPs & \# NCIs & STY & PPV & MCC & \multicolumn{1}{c|}{Structure complexity} \\ \hline \hline
\textbf{2F8K} & \textbf{16} & \textbf{6} & \textbf{14} & \textbf{85} & \textbf{85} & \textbf{0.8571} & \textbf{Hairpin} \\ \hline
\textbf{2AB4} & \textbf{20} & \textbf{6} & \textbf{20} & \textbf{100} & \textbf{90} & \textbf{0.9534} & \textbf{Hairpin} \\ \hline
\textbf{361D} & \textbf{20} & \textbf{5} & \textbf{17} & \textbf{70} & \textbf{57} & \textbf{0.6351} & \textbf{Hairpin} \\ \hline
2ANN & 23 & 3 & 24 & 75 & 66 & 0.7071 & Hairpin \\ \hline
1RLG & 25 & 5 & 22 & 95 & 63 & 0.7793 & Hairpin, internal loop \\ \hline
2QUX & 25 & 9 & 22 & 90 & 71 & 0.8058 & Hairpin \\ \hline
387D & 26 & 4 & 23 & 86 & 68 & 0.7744 & Hairpin \\ \hline
1MSY & 27 & 6 & 39 & 97 & 92 & 0.9502 & Hairpin \\ \hline
1L2X & 28 & 8 & 34 & 88 & 88 & 0.8823 & Pseudoknot \\ \hline
2AP5 & 28 & 8 & 29 & 82 & 66 & 0.7427 & Pseudoknot \\ \hline
1JID & 29 & 8 & 31 & 93 & 72 & 0.8235 & Hairpin, internal loop \\ \hline
1OOA & 29 & 8 & 29 & 93 & 72 & 0.8242 & Hairpin, internal loop \\ \hline
430D & 29 & 6 & 37 & 94 & 77 & 0.8577 & Hairpin, internal loop \\ \hline
3SNP & 30 & 12 & 31 & 93 & 85 & 0.8932 & Hairpin, internal loop \\ \hline
2OZB & 33 & 10 & 33 & 93 & 79 & 0.8641 & Hairpin, internal loop \\ \hline
1MJI & 34 & 10 & 44 & 84 & 84 & 0.8409 & Hairpin, internal loop \\ \hline
1ET4 & 35 & 8 & 40 & 67 & 84 & 0.7546 & Pseudoknot \\ \hline
2HW8 & 36 & 12 & 44 & 93 & 80 & 0.8655 & Hairpin, internal loop \\ \hline
1I6U & 37 & 15 & 47 & 91 & 89 & 0.9053 & Hairpin, internal loop \\ \hline
1F1T & 38 & 10 & 38 & 81 & 63 & 0.7184 & Hairpin, internal loop \\ \hline
\textbf{1ZHO} & \textbf{38} & \textbf{13} & \textbf{46} & \textbf{95} & \textbf{83} & \textbf{0.8911} & \textbf{Hairpin, internal loop} \\ \hline
1S03 & 47 & 18 & 53 & 88 & 79 & 0.8404 & Hairpin, internal loop \\ \hline
1XJR & 47 & 15 & 55 & 83 & 80 & 0.8215 & Hairpin, internal loop \\ \hline
1U63 & 49 & 17 & 50 & 94 & 65 & 0.7833 & Hairpin, internal loop \\ \hline
2PXB & 49 & 16 & 66 & 98 & 94 & 0.9632 & Hairpin, internal loop \\ \hline
2FK6 & 53 & 20 & 58 & 77 & 70 & 0.7385 & Pseudoknot, 3-way junction \\ \hline
3E5C & 53 & 21 & 65 & 84 & 73 & 0.7877 & 3-way junction (riboswitch) \\ \hline
1MZP & 55 & 17 & 73 & 64 & 73 & 0.6876 & Hairpin internal \\ \hline
1DK1 & 57 & 24 & 65 & 100 & 89 & 0.9436 & 3-way junction \\ \hline
1MMS & 58 & 20 & 86 & 74 & 82 & 0.7814 & 3-way junction \\ \hline
3EGZ & 65 & 23 & 72 & 70 & 66 & 0.6849 & 3-way junction (riboswitch) \\ \hline
2QUS & 69 & 26 & 81 & 75 & 76 & 0.7577 & Pseudoknot, 3-way junction \\ \hline
1KXK & 70 & 28 & 87 & 96 & 92 & 0.9440 & Hairpin, internal loop \\ \hline
2DU3 & 71 & 27 & 75 & 78 & 70 & 0.7433 & 4-way junction (tRNA) \\ \hline
\textbf{2OIU} & \textbf{71} & \textbf{29} & \textbf{84} & \textbf{90} & \textbf{83} & \textbf{0.8692} & \textbf{3-way junction (riboswitch)} \\ \hline
1SJ4 & 73 & 19 & 83 & 78 & 81 & 0.7976 & Pseudoknot, 4-way junction \\ \hline
1P5O & 77 & 29 & 86 & 97 & 77 & 0.8716 & Hairpin, internal loop \\ \hline
3D2G & 77 & 28 & 103 & 80 & 88 & 0.8435 & 3-way junction (riboswitch) \\ \hline
2HOJ & 79 & 27 & 100 & 87 & 84 & 0.8572 & 3-way junction (riboswitch) \\ \hline
\textbf{2GDI} & \textbf{80} & \textbf{32} & \textbf{100} & \textbf{84} & \textbf{80} & \textbf{0.8197} & \textbf{3-way junction (riboswitch)} \\ \hline
2GIS & 94 & 36 & 125 & 87  & 82 & 0.8485 & Pseudoknot, 4-way junction (riboswitch) \\ \hline
1LNG & 97 & 38 & 124 & 85 & 79 & 0.8254 & 3-way junction (SRP) \\ \hline
\textbf{1MFQ} & \textbf{128} & \textbf{49} & \textbf{164} & \textbf{81} & \textbf{76} & \textbf{0.7895} & \textbf{3-way junction (SRP)} \\ \hline
\end{tabular}
\end{table}

Given the recent progress made in RNA secondary structure prediction \cite{LaingAndSchlick2010}, \cite{ReinharzEtAl2013}, we believe that canonical base pairs may be routinely predicted with a fair accuracy. Therefore, we have allowed the program BkTree to accept known or predicted canonical base pairs along with the query sequence as input. Note that the knowledge of canonical base pairs does not necessarily imply the whole secondary structure, which is often a part of input to most of the existing RNA 3D prediction methods. In our test, we extracted canonical base pairs of a RNA from FR3D analyzed interactions \cite{SarverEtAl2008}. 

\subsection{Overall Performance}

We evaluated the quality of the predicted nucleotide interactions by the sensitivity (STY) and positive predictive value (PPV) against the FR3D-analyzed interactions \cite{SarverEtAl2008}. In order to take into account the effects of both true positive and false positive rates in one measure, the \textit{Matthews correlation coefficient} (MCC), defined in \cite{LaingAndSchlick2010} as MCC $:= \sqrt{\mathrm{PPV}\times \mathrm{STY}}$, was also calculated. 


Table \ref{withBbsTable} summarizes the overall performance of BkTree on the benchmark set. On a large majority of RNAs, the sensitivity is decently high. Note that the STY and PPV calculations excluded the canonical base pairs.The sensitivity result indicates that our method has a high accuracy in identifying non-canonical interactions that may be crucial to tertiary structures. This is true even for those longer RNAs. We further note that for the 7 RNAs that were not included in the training data, BkTree also performed extremely well.

\subsection{Performance Comparison with Other Methods}
We compared our program BkTree with the programs MC, Rosetta, and NAST on the capability to predict nucleotide interactions. These other methods had been surveyed and evaluated in \cite{LaingAndSchlick2010} based on their ability to identify both base pairing and base stacking interactions. We removed  base-phosphate and base-ribose interactions from our prediction results. We incorporated the canonical base pairs into our results because these other methods include all interactions from the input secondary structure. 

\begin{figure}[!ht]
\centering
\includegraphics[scale=0.38]{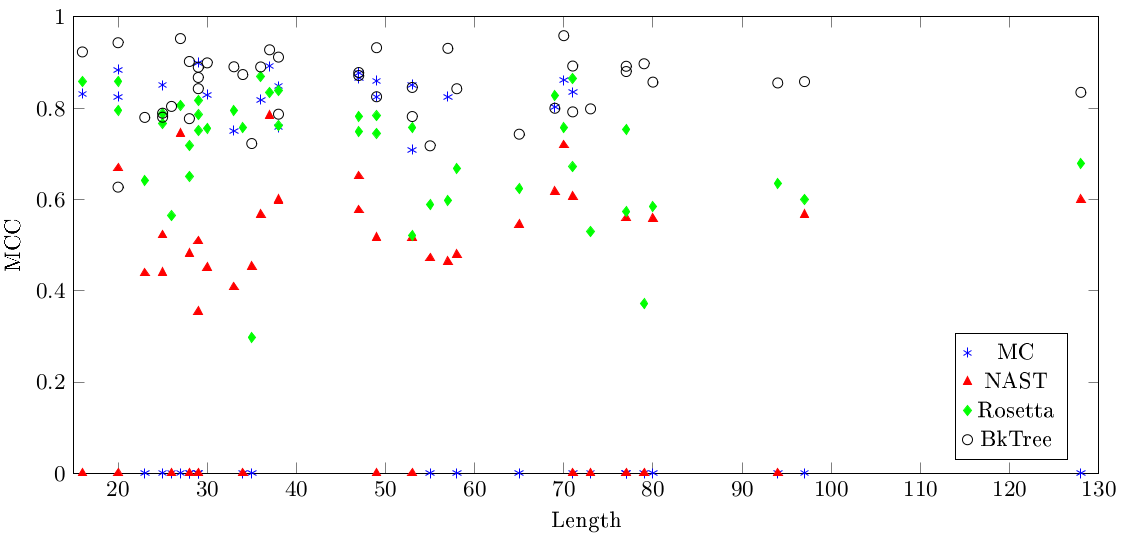}
\caption{Comparison of the MCC generated by MC, NAST, Rosetta and BkTree. The MCC  of the 43 RNAs are calculated by including canonical base pairs in the results and sorted by their lengths. The plot was derived by merging the results obtained by BkTree and the data computed in the survey \cite{LaingPersonal2014,LaingAndSchlick2010}. In that survey, the tertiary structure predictions with the other 3 methods were based on resolved secondary structures and the secondary structures were included in the calculations. Therefore, the canonical base pairs were also been added to the perdiction results by BkTree. }
\label{laingMCCCompare}
\end{figure}

Figure~\ref{laingMCCCompare} shows the MCC curves for MC, Rosetta, NAST, and BkTree on the benchmark set of RNAs. Data of RNAs failed by a program were not included in the calculation. We note that for every RNA, these other programs produced more than one conformation so the results were averaged for these comparisons. The figure demonstrates that BkTree overall outperformed the other three programs in predicting non-canonical base pairing and base stacking interactions. 

Table~\ref{avg4methods} gives comparisons on average performance between the four methods. In general, Bktree produced much better average results than Rosetta and NAST, and comparable average results with MC, for which  BkTree shows better average STY value than MC, whereas MC gives better average PPV. On MCC values, BkTree had an edge over MC. On RNAs of length $\geq 50$, BkTree maintained almost the same average MCC as it did on the whole set. 


\begin{table}[!ht]
\centering
\caption{Average performances of MC, Rosetta, NAST and BkTree, with results in two categories: average over all successfully resolved RNAs and average over all successfully resolved RNAs of length $>50$.  The best performance data are displayed in bold. }
\begin{tabular}{c|c|c|c|c|c|c|c|c|}
\cline{2-9}
 & \multicolumn{4}{c|}{All RNAs} & \multicolumn{4}{c|}{RNAs of length \textgreater50} \\ \cline{2-9} 
 & Success/Total & STY & PPV & MCC & Success/Total & STY & PPV & MCC \\ \hline \hline
\multicolumn{1}{|c|}{MC} & 21/43 & 80.7 & \textbf{86.2} & 0.8344 & 6/18 & 77.1 & \textbf{86.0} & 0.8145 \\ \hline
\multicolumn{1}{|c|}{Rosetta} & 43/43 & 62.8 & 80.3 & 0.7101 & 18/18 & 53.4 & 78.5 & 0.6474 \\ \hline
\multicolumn{1}{|c|}{NAST} & 30/43 & 44.5 & 68.2 & 0.5508 & 12/18 & 44.0 & 71.4 & 0.5604 \\ \hline
\multicolumn{1}{|c|}{BkTree} & 43/43 & \textbf{88.6} & 81.3 & \textbf{0.8482} & 18/18 & \textbf{86.0} & 82.7 & \textbf{0.8433} \\ \hline
\end{tabular}
\label{avg4methods}
\end{table}

\subsection{Significance to 3D conformation prediction}\label{3dConfSection}

To evaluate the significance of our method to 3D conformation prediction, we used MC-Sym \cite{ParisienAndMajor2008} to model 3D conformations from the interactions predicted by BkTree and calculated RMSDs against the resolved structures. We note that MC-Sym does not accept interactions of categories other than base-pair and base stacking; the correctly predicted base-phosphate and base-ribose interactions by our methods were discarded by MC-Sym to produce 3D folds. The \textit{deviation index} (DI) \cite{ParisienAndCruz2009}, a measure that accounts for both RMSD and MCC, defined as the quotient of them, was also calculated. Table \ref{compare_Laing} presents the performance values on the 4 representative RNAs chosen in \cite{LaingAndSchlick2010} which typically contain two hairpins and two junctions. Since both MC and Rosetta allow prediction of multiple optimal or suboptimal folds, we chose the averaged values of their solutions. We note that to model 3D conformations with MC using our predicted interaction data, we needed the secondary structure of the tested  RNA to be covered by the input canonical base pairs together with the interactions predicted by BkTree. RNA 2QUS failed on this requirement. The averaged RMSDs achieved by BkTree for the rest 3 RNAs are significantly smaller than those achieved by MC and Rosetta.

\begin{table}[!ht]
\centering
\caption{List of performance values predicted using MC, Rosetta and BkTree on 4 representative RNAs chosen by \cite{LaingAndSchlick2010}. The results generated MC and Rosetta are obtained from the survey paper \cite{LaingPersonal2014,LaingAndSchlick2010}. For every RNA, the best results are displayed in bold.}
\begin{tabular}{cc|c|c|c|c|c|c|c|c|c|c|c|c|c|c|c|}
\cline{3-17}
 &  & \multicolumn{5}{c|}{MC} & \multicolumn{5}{c|}{Rosetta} & \multicolumn{5}{c|}{BkTree} \\ \hline
\multicolumn{1}{|c|}{PDB} & Length & STY  & PPV & MCC & RMSD & DI & STY & PPV & MCC & RMSD & DI & STY & PPV & MCC & RMSD & DI \\ \hline \hline
\multicolumn{1}{|c|}{1KXK} & 70 & 81 & 89 & 0.849 & 9.49 & 11.16 & 74 & 85 & 0.793 & 17.23 & 21.69 & 97 & 94 & \textbf{0.9589} &  \textbf{8.33} & \textbf{8.68} \\ \hline
\multicolumn{1}{|c|}{1XJR} & 47 & 76 & 87 & 0.8131 & 8.74 & 10.74 & 71 & 83 & 0.7676 & 11.63 & 15.21 & 91 & 84 & \textbf{0.8782} & \textbf{6.00} & \textbf{6.83} \\ \hline
\multicolumn{1}{|c|}{2OIU} & 71 & 76 & 92 & 0.8361 & 16.85 & 20.14 & 63 & 87 & 0.7403 & 18.10 & 24.72 & 92 & 86 & \textbf{0.8925} & \textbf{13.21} & \textbf{14.8} \\ \hline
\multicolumn{1}{|c|}{2QUS} & 69 & 78 & 86 & \textbf{0.819} & 18.41 & 22.44 & 58 & 86 & 0.7062 & 15.73 & 22.80 & 80 & 80 & 0.8 & - & - \\ \hline
\end{tabular}
\label{compare_Laing}
\end{table}

\section{Discussion and Conclusion}\label{disSection}
Our method is the first to {\it ab initio} predict RNA non-canonical interactions of all types. Evaluation of the results have highlighted its potential as an important step toward accurate {\it ab initio} 3D structure prediction. We attribute the encouraging preliminary results to the recent growth of knowledge in high-resolution nucleotide interaction data as well as to the novel backbone $k$-tree modeling of nucleotide interaction relationships. The latter makes it possible to markedly reduce the space of solutions for the nucleotide interaction prediction problem to one that can be feasibly searched in polynomial time.

Our method differs from others also in its direct prediction of nucleotide interactions whereas the others mostly attempt 3D conformation construction before producing nucleotide interactions. The difference makes it difficult to compare their performances, especially when a 3D structure is not the direct output of a software, e.g., RNA-MoIP \cite{ReinharzEtAl2013}. Therefore, the MCC comparison with MC was probably more appropriate than the comparison with RNA-MoIP, since the results of MC were based on interactions from the RNA Structure Atlas and so did BkTree, while RNA-MoIP used Interaction Network Fidelity \cite{GendronEtAl2001} in calculating the MCC values. The contrast is more evident when using MC-Sym to model 3D conformations from interaction data predicted by BkTree. Even though the predicted base-phosphate and base-ribose interactions have to be discarded, the resulted RMSDs seem to correlate with the MCC values (Table~\ref{compare_Laing}). 

The evaluation tests have also revealed some issues with BkTree. First, the complexity of structures has an impact on our prediction results. Typically, BkTree underperformed on some of the RNAs with pseudoknots or 4-way junctions. Table~\ref{compare_Laing} shows that BkTree loses to MC on MCC value for only one representative RNA 2QUS, which contains a pseudoknot. The underperformance is likely due to the 3-tree model that is a little too weak for complex structures.  For example, the best 3-tree can include at most 83 interactions out of total 95 interactions of tRNA 2DU3, indicating a higher treewidth is needed for the NIR graph of this RNA. To improve prediction performance for such RNAs, an algorithm may need to be based on the backbone $4$-tree model. Our method is not ineffective for handling multi-way junctions or pseudoknots, e.g., RNA 2GIS in Table \ref{withBbsTable}. Fixing a specific $k$-tree model, it is the NIR graph treewidth of an RNA that determines the performance on the RNA.

Second, the NIR graph treewidth is also related to scalability of our method. The current algorithm for the nucleotide prediction problem has the complexity $O(n^3)$ for both time and memory requirements. With a large hidden constant in the polynomial, the implemented program BkTree typically runs in 2 to 3 hours on an RNA of length 100 and uses several Gigabytes of memory. This is because the current prototype has aimed at accuracy without optimization in computational efficiency. However, the problem (based on the $k$-tree model) has an inherent complexity of $O(n^k)$; our method is scalable to suit longer and more complex RNAs, e.g., which require the $4$-tree model.

Third, due to the lack of tools to model 3D conformations from nucleotide interactions of all types, it is an immediate future task of ours is to develop such a tool that can be pipelined with a program like  BkTree for {\it ab initio} 3D structure prediction. We perceive such a task to be feasible. This is because the output of program BkTree contains not only the predicted nucleotide interactions but also a backbone  $3$-tree that decomposes nucleotides according to their interconnectivity. The given 3-tree can be the basis for very efficient algorithms for computing a desirable optimization function on 3D conformations \cite{ArnborgAndProskurowski1989}.









\section*{Acknowledgments}
We thank Christian Laing for the provided raw data used in the survey paper \cite{LaingAndSchlick2010}. This work was supported in part by NSF IIS grant (award No: 0916250).


\end{document}